\documentstyle[12pt]{article}\textheight 230mm\textwidth 160mm
\newcommand{\bi}{\bibitem}
\newcommand{\be}{\begin{eqnarray}}
\newcommand{\ee}{\end{eqnarray}}
\newcommand{\nn}{\nonumber}

\begin{document}
\hspace*{11cm}\vspace{-2mm}MPI-PhT/94-71\\
\hspace*{11.6cm}\vspace{-2mm}KANAZAWA-94-23\\
\hspace*{11.6cm}\vspace{-2mm}HD-THEP-94-40\\
\hspace*{11.6cm}November 1994
\begin{center}
{\Large\bf Gauge-Yukawa Unification:\\
Going Beyond GUTs$^{\dag}$}
\end{center}

\vspace*{0.1cm}
\begin{center}{\sc Jisuke Kubo}$\ ^{(1)}$,
{\sc Myriam
Mondrag{\' o}n}$\ ^{(2)}$\vspace{-1mm} and \\
{\sc George Zoupanos}$\ ^{(3),*}$
\end{center}
\begin{center}
{\em $\ ^{(1)}$ Max-Planck-Institut f\"ur Physik,
 Werner-Heisenberg-Institut \vspace{-2mm}\\
D-80805 Munich, Germany}\vspace{-3mm} \\and\vspace{-3mm} \\
{\em College of Liberal Arts, Kanazawa University,
920-11 Kanazawa, Japan}\\
{\em $\ ^{(2)}$ Institut f{\" u}r Theoretische Physik,
Philosophenweg 16\vspace{-2mm}\\
D-69120 Heidelberg, Germany}\\
{\em $\ ^{(3)}$ Physics Department, National Technical\vspace{-2mm}
University\\ GR-157 80 Zografou, Athens, Greece }  \end{center}
{\sc\large Abstract}

\noindent
We discuss the basic idea of the Gauge-Yukawa Unification that is based
on the principle of the reduction of couplings.
This method of unification  relies on the search of
successful renormalization group invariant relations among couplings,
which do not originate from
 symmetry principles. The predictive power of Grand Unified
Theories can be increased by this method, predicting for
instance  values of the top quark mass consistent with the
recent experimental data. The hope is that this unification attempt
might shed further light on the origin of the Yukawa sector of the
standard model.
\vspace*{1cm}
\footnoterule
\vspace*{2mm}
\noindent
$^{*}$Partially supported by C.E.U. projects
(SC1-CT91-0729; CHRX-CT93-0319).\\
$ ^{\dag}$ Presented by J. Kubo at the  \vspace{-2mm}
{\em IX Workshop on High
Energy Physics and Quantum Field Theory}, Zvenigord,
Moscow Region, 16-22 September 1994, to appear in the
proceedings.

\newpage
\pagestyle{plain}
\section{Introduction}
The success of the standard model \cite{dydak,schild} shows that we have
at hand a highly nontrivial part of a more fundamental theory
of elementary particle physics, which challenges theorists
to understand at least some of the plethora of its free
parameters.

In constructing realistic field theory models,
their renormalizability
has played undoubtedly an important r{\^ o}le  \cite{veltman}.
In particular,
the structure of the independent parameters in a given theory
is basically fixed by its renormalizability.
Therefore, the traditional recipe of reducing the number
of the independent parameters was to
impose symmetries that are compatible
 with  renormalizability.
Grand Unified Theories
(GUTs) \cite{georgi1,fritzsch1}
 relate in this way not only the gauge couplings of the standard model,
but also its Yukawa
couplings. In fact, the
Georgi-Glashow $SU(5)$ model	\cite{georgi1} was very successful
in qualitatively predicting the value of the $\sin^{2} \theta_{\rm W}$
as well as the mass
ratio $m_{\tau}/m_{b}$ \cite{buras1}.

A logical extension of the GUT idea is to attempt to relate the couplings
of the  gauge and Yukawa sectors, which we would like to call Gauge-Yukawa
Unification. However, within the framework of field theory
(assuming that all the particles appearing in a
theory are elementary), the extended supersymmetry \cite{fayet1}
is the only symmetry that
could be used to achieve a Gauge-Yukawa Unification (GYU).
Unfortunately, theories based on extended supersymmetries
seem to introduce more
serious and difficult phenomenological problems to
be solved than those of the standard model \cite{aguila}.

There exists an alternative way to
unify couplings
which is based on the fact that
within the framework of
renormalizable field theory, one can find renormalization group
invariant (RGI) relations among parameters
and improve in this way the calculability
and predictive power of a given
theory \cite{zimmermann1}-\cite{kubo3}
(see also ref. \cite{scardron} for an alternative method).
Let us briefly describe this idea below.

Any RGI relation among couplings
(which does not depend on the renormalization
scale $\mu$ explicitly) can be expressed
in the implicit form as $\Phi (g_1,\cdots,g_N) ~=~\mbox{const.}$,
where $\Phi $ has to
satisfy the partial differential equation (PDE):
\be
\mu\,\frac{d \Phi}{d \mu} &=& {\vec \nabla}\cdot {\vec \beta} ~=~
\sum_{i=1}^{N}
\,\beta_{i}\,\frac{\partial \Phi}{\partial g_{i}}~=~0~,
\ee
where $\beta_i$ is the $\beta$-function of $g_i$.
This PDE is equivalent
to the set of the ordinary differential equations,
the so-called reduction equations (REs)\cite{zimmermann1},
\be
\beta_{g} \,\frac{d g_{i}}{d g} &=&\beta_{i}~,~i=1,\cdots,A~,
\ee
where $g$ and $\beta_{g}$ are the primary
coupling and its $\beta$-function,
and the count on $i$ does not include it.
Since maximally ($N-1$) independent
RGI relations
in the $N$-dimensional space of couplings
can be imposed by $\Phi_i$'s, one could in principle
express all the couplings in terms of
a single coupling $g$.
 The strongest requirement is to demand
 power series solutions to the REs,
\be
g_{i} &=& \sum_{n=0}\,r_{i}^{(n+1)}\,g^{2n+1}~,
\ee
which formally preserve perturbative renormalizability, where
$r_{i}^{(n)}$'s are the expansion coefficients.
The possibility of this coupling unification
is without any doubt very
attractive because a ``completely reduced'' theory would contain
only one independent coupling $g$. However this ideal case can be
unrealistic. Therefore, one often is lead to impose fewer RGI
constraints in order to preserve a given theory in a realistic
framework, and to introduce the idea of partial reduction \cite{kubo1}.
Among the existing possibilities
in the framework of supersymmetric $SU(5)$ GUTs,
there are two models that are singled out by being strongly
motivated \cite{kapet1,kubo2}. The first one is the $SU(5)$-Finite
Unified Theory (FUT) (see refs. \cite{kazakov}, and \cite{kapet1} and
references therein).
In this theory  \cite{kapet1} , there exist
RGI relations among gauge and Yukawa couplings that yield
the vanishing of all $\beta$-functions
to all orders in perturbation theory \cite{sibold1}.
 The second is the minimal supersymmetric $SU(5)$
model \cite{dgs} which can be successfully
partially-reduced \cite{kubo2}.
The latter  is
attractive because of its simplicity.

Clearly, in both cases the existence of
a covering GUT is assumed so that the unification
of the gauge couplings of the standard model
is of a group theoretic nature. In ref. \cite{kubo3},
we have
examined the power of the RGI method by considering
theories without covering GUTs, and
found
that the  supersymmetrized model based on the  Pati-Salam gauge
group is phenomenologically viable.
The predictability of the model
on the known physics is improved by the
present Gauge-Yukawa unification method and, in particular,
the model contains
only one gauge coupling instead of three.

\section{The Principle of the Partial Reduction of Couplings}
Here we would like to briefly outline the basic tool
of the partial reduction
(see refs. \cite{kubo1,kubo2} for details) which was mentioned above.
For many cases, it is convenient
to work with the absolute square of $g_{i}$'s, and therefore we
 define the tilde couplings by
\be
\tilde{\alpha}_{i} &\equiv&
\frac{\alpha_{i}}{\alpha}~,~i=1,\cdots,N~,
\ee
where
$ \alpha ~=~|g|^2/4\pi$ and $\alpha_{i} ~=~
|g_{i}|^2/4\pi$ ($i$ does not include the primary
coupling). We assume that their evolution equations take the form
\be
\frac{d \alpha}{dt} &=&-b^{(1)}\,\alpha^2+\cdots~,\nn\\
\frac{d\alpha_i}{dt} &=&-b^{(1)}_{i}\,\alpha_{i}\alpha+
\sum_{j,k}b^{(1)}_{i,jk}\,\,\alpha_j\alpha_k+\cdots~,
\ee
in perturbation theory.

We eliminate $t$ and derive the evolution equations for the
tilde couplings
\be
\alpha \frac{d \tilde{\alpha}_{i}}{d\alpha} &=&
(\,-1+\frac{b^{(1)}_{i}}{b^{(1)}}\,)\, \tilde{\alpha}_{i}
-\sum_{j,k}\,\frac{b^{(1)}_{i,jk}}{b^{(1)}}
\,\tilde{\alpha}_{j}\, \tilde{\alpha}_{k}+\sum_{r=2}\,
(\frac{\alpha}{\pi})^{r-1}\,\tilde{b}^{(r)}_{i}(\tilde{\alpha})~,
\ee
where
$\tilde{b}^{(r)}_{i}(\tilde{\alpha})~(r=2,\cdots)$
are power series of $\tilde{\alpha}_{i}$ and can be computed
from the $r$-th loop $\beta$-functions.
To proceed, we solve the set of the algebraic equations
\be
(\,-1+\frac{b^{(1)}_{i}}{b^{(1)}}\,)\, \rho_{i}^{(1)}
-\sum_{j,k}\frac{b^{(1)}_{i,jk}}{b^{(1)}}
\,\rho_{j}^{(1)}\, \rho_{k}^{(1)}&=&0~,
\ee
and assume that their solutions $\rho_{i}^{(1)}$'s have the form
\be
\rho_{i}^{(1)}&=&0~\mbox{for}~ i=1,\cdots,N'~;~
\rho_{i}^{(1)} ~>0 ~\mbox{for}~i=N'+1,\cdots,N~.
\ee
Given the set of the solutions above,
we regard $\tilde{\alpha}_{i}$ with $i \leq N'$
 as small
perturbations  to the
undisturbed system which is defined by setting
$\tilde{\alpha}_{i}$, with $i \leq N'$, equal to zero.
It is possible \cite{zimmermann1}
to verify, at the one-loop level
the existence of
the unique power series solution
\be
\tilde{\alpha}_{i}&=&\rho_{i}^{(1)}+\sum_{r=2}\rho^{(r)}_{i}\,
(\frac{\alpha}{\pi})^{r-1}~,~i=N'+1,\cdots,N~
\ee
of the reduction equations (6) to all orders
in  the undisturbed system.
These are RGI relations among couplings and keep formally
the perturbative renormalizability of the undisturbed system.
So in the undisturbed system there is only {\em one independent}
coupling, the primary coupling $\alpha$.

 The small
 perturbations caused by nonvanishing $\tilde{\alpha}_{i}$,
 with $i \leq N'$,
enter in such a way that the reduced couplings,
i.e., $\tilde{\alpha}_{i}$  with $i > N'$,
become functions not only of
$\alpha$ but also of $\tilde{\alpha}_{i}$
 with $i \leq N'$.
It turned out that, to investigate such partially
reduced systems, it is most convenient to work with the partial
differential equations
\be
\{~~\tilde{\beta}\,\frac{\partial}{\partial\alpha}
+\sum_{a=1}^{N'}\,
\tilde{\beta_{a}}\,\frac{\partial}{\partial\tilde{\alpha}_{a}}~~\}~
\tilde{\alpha}_{i}(\alpha,\tilde{\alpha})
&=&\tilde{\beta}_{i}(\alpha,\tilde{\alpha})~,\\
\tilde{\beta}_{i(a)}~=~\frac{\beta_{i(a)}}{\alpha^2}
-\frac{\beta}{\alpha^{2}}~\tilde{\alpha}_{i(a)}
&,&
\tilde{\beta}~\equiv~\frac{\beta}{\alpha}~,\nn
\ee
 which are equivalent
to the reduction equations (6) (we let
$a,b$ run from $1$ to $N'$ and $i,j$ from $N'+1$ to $N$,
in order to avoid confusion).
We then look for solutions of the form  \cite{kubo1,kubo2}
\be
\tilde{\alpha}_{i}&=&\rho_{i}^{(1)}+
\sum_{r=1}\,(\frac{\alpha}{\pi})^{r-1}\,f^{(r)}_{i}
(\tilde{\alpha}_{a})~,~i=N'+1,\cdots,N~,
\ee
where $ f^{(r)}_{i}(\tilde{\alpha}_{a})$ are supposed to be
power series of
$\tilde{\alpha}_{a}$. This particular type of solution
can be motivated by requiring that, in the limit of vanishing
perturbations, we obtain the undisturbed
solutions (9) \cite{kubo1,zimmermann3}, i.e.,
$f_{i}^{(1)}(0)=0~,~ f_{i}^{(r)}(0)=\rho_{i}~\mbox{for}~r \geq 2$.
Again it is possible to obtain  the sufficient conditions for
the uniqueness of $ f^{(r)}_{i}$ in terms of the lowest order
coefficients. Thus, the partially-reduced system contains
the primary coupling $\alpha$ and the disturbing
ones $\tilde{\alpha}_{a}$'s only, thereby increasing
the predictive power of the original system.

\section{An Example}
In the traditional GUT scheme, there exists
a covering GUT so that the unification
of the gauge couplings of the standard model
is of a group theoretic nature. Here
we would like to
examine the power of the RGI method by considering
theories without covering GUTs \cite{kubo3}.
Obviously, in order the RGI method for the gauge coupling
unification to  work,
the gauge couplings should
have the same asymptotic behavior.
Note that this common behavior is absent
in the standard model with three families.
A way to achieve a common asymptotic behavior of all the
different gauge couplings is to embed
$SU(3)_{C}\times SU(2)_{L}\times U(1)_{Y}$ to some
non-abelian gauge group, and so
we introduce  new physics
at a very high energy scale and increase
the predictability of the model
on the known physics by unifying the gauge and part of Yukawa sectors
on the basis of the reduction principle. We \cite{kubo3} have found
that the minimal phenomenologically viable model is based on the gauge
group of Pati and
Salam \cite{pati1}-- ${\cal G}_{\rm PS}\equiv
SU(4)\times SU(2)_{R}\times
SU(2)_{L}$.
We recall that $N=1$ supersymmetric  models based on this
gauge group have been studied with renewed interest because they could
in principle be derived from superstrings \cite{anton1}.

In our supersymmetric, Gauge-Yukawa unified model
based on $ {\cal G}_{\rm PS}$ \cite{kubo3}, three generations of
quarks and leptons  are accommodated by six chiral supermultiplets, three
in $({\bf 4},{\bf 2},{\bf 1})$ and three  $({\bf \overline{4}},{\bf
1},{\bf 2})$, which we denote by $\Psi^{(I)\mu~ i_R}$ and $
\overline{\Psi}_{\mu}^{(I) i_L}$ ($I$ runs over the three generations,
and
$\mu,\nu~(=1,2,3,4)$ are the $SU(4)$ indices while
$i_R~,~i_L~(=1,2)$
stand for the
$SU(2)_{L,R}$ indices).
The Higgs supermultiplets
in $({\bf 4},{\bf 2},{\bf 1})$,
$({\bf \overline{4}},{\bf 2},{\bf 1})$
and  $({\bf 15},{\bf 1},{\bf 1})$ are denoted by
$ H^{\mu ~i_R}~,~
\overline{H}_{\mu ~i_R} $, and $\Sigma^{\mu}_{\nu}$ respectively. They
 are responsible for the spontaneous
symmetry breaking (SSB) of $SU(4)\times SU(2)_{R}$ down
to $SU(3)_{C}\times U(1)_{Y}$.
The SSB of $U(1)_{Y}\times
SU(2)_{L}$ is then achieved by the nonzero VEV of
$h_{i_R i_L}$ which is in $({\bf 1},{\bf 2},{\bf 2})$. In addition to
these Higgs supermultiplets, we introduce $G^{\mu}_{\nu~i_R i_L}~
({\bf 15},{\bf 2},{\bf 2})~,
{}~\phi~({\bf 1},{\bf 1},{\bf 1})$ and
$\Sigma^{' \mu}_{\nu}~({\bf 15},{\bf 1},{\bf 1})$.
The $G^{\mu}_{\nu~i_R i_L}$ is introduced to realize
the $SU(4)\times SU(2)_{R}\times
SU(2)_{L}$ version of the Georgi-Jarlskog type
ansatz \cite{georgi4} for
the mass matrix of leptons and quarks while $\phi$
is supposed to mix with the right-handed neutrino
supermultiplets at a high energy scale.
With these in mind, we write down
the superpotential $W$ of the model,
which is the sum of the following terms:
\be
W_{Y} &=&\sum_{I,J=1}^{3}g_{IJ}\,\overline{\Psi}^{(I) i_R}_{\mu}
\,\Psi^{(J)\mu~ i_L}~h_{i_R i_L}~,~
W_{GJ} ~=~g_{GJ}\,
\overline{\Psi}^{(2)i_R}_{\mu}\,
G^{\mu}_{\nu~i_R j_L}\,\Psi^{(2)\nu~ j_L}~,\nn\\
W_{NM} &=&
\sum_{I=1,2,3}\,g_{I\phi}~\epsilon_{i_R j_R}\,\overline{\Psi}^{(I)
i_R}_{\mu} ~H^{\mu ~j_R}\,\phi~,\\
W_{SB} &=&
g_{H}\,\overline{H}_{\mu~ i_R}\,
\Sigma^{\mu}_{\nu}\,H^{\nu ~i_R}+\frac{g_{\Sigma}}{3}\,
\mbox{Tr}~[~\Sigma^3~]+
\frac{g_{\Sigma '}}{2}\,\mbox{Tr}~[~(\Sigma ')^2\,\Sigma~]~,\nn\\
W_{TDS} &=&
\frac{g_{G}}{2}\,\epsilon^{i_R j_R}\epsilon^{i_L j_L}\,\mbox{Tr}~
[~G_{i_R i_L}\,
\Sigma\,G_{j_R j_L}~]~,\nn\\
W_{M}&=&m_{h}\,h^2+m_{G}\,G^2+m_{\phi}\,
\phi^2+m_{H}\,\overline{H}\,H+
m_{\Sigma}\,\Sigma^2+
m_{\Sigma '}\,(\Sigma ')^2~.\nn
\ee
Although $W$ has the parity, $\phi\to -\phi$
and $\Sigma ' \to -\Sigma '$,
it is not the most general potential, and,
by virtue of the non-renormalization theorem,
this does not contradict the philosophy of
the coupling unification by the RGI method.

We denote the gauge couplings of $SU(4)\times SU(2)_{R}\times
SU(2)_{L}$
by $\alpha_{4}~,~\alpha_{2R}$, and $\alpha_{2L}$
respectively. The gauge coupling for $U(1)_{Y}$, $\alpha_1$, normalized
in the usual GUT inspired manner, is given by
$1/\alpha_{1} ~=~2/5\alpha_{4}+
3/5 \alpha_{2R}~$.
In principle, the primary coupling can be any one of the couplings.
But it is more convenient to choose a gauge coupling as the primary
one because the one-loop $\beta$ functions for a gauge coupling
depends only on its own gauge coupling. For the present model,
we use $\alpha_{2L}$ as the primary one.
Since the gauge sector for the one-loop $\beta$ functions is closed,
the solutions of the fixed point equations (7) are
independent on the Yukawa and Higgs couplings. One easily obtains
$
\rho_{4}^{(1)} =8/9~,~\rho_{2R}^{(1)}~=~4/5$,
so that the
RGI relations (11) at the one-loop level become
\be
\tilde{\alpha}_{4} &=&\frac{\alpha_4}{\alpha_{2L}}~=~
\frac{8}{9}~,~\tilde{\alpha}_{1} ~=~\frac{\alpha_1}{\alpha_{2L}}~=~
\frac{5}{6}~.
\ee

The solutions in the Yukawa-Higgs sector strongly
depend on the result of the gauge sector. After slightly involved
algebraic computations, one finds that
most predictive solutions contain at least
three vanishing $\rho_{i}^{(1)}$'s.
Out of these solutions, there are two that
 exhibit the most predictive
power and moreover they satisfy
the neutrino mass relation
$m_{\nu_{\tau}}~>~m_{\nu_{\mu}}~,~
m_{\nu_{e}}$.
For the first solution we have $\rho_{1\phi}^{(1)}=
\rho_{2\phi}^{(1)}=
\rho_{\Sigma}^{(1)}=0$, while for the second one,
$  \rho_{1\phi}^{(1)}=
\rho_{2\phi}^{(1)}=
\rho_{G}^{(1)}=0 $.
One then finds that for these two cases  the power series solutions
(11) take the form
\be
\tilde{\alpha}_{GJ} &\simeq &
\left\{
\begin{array}{l} 1.67 - 0.05 \tilde{\alpha}_{1\phi}
+
0.004 \tilde{\alpha}_{2\phi}
 - 0.90\tilde{\alpha}_{\Sigma}+\cdots \\
 2.20 - 0.08 \tilde{\alpha}_{2\phi}
 - 0.05\tilde{\alpha}_{G}+\cdots
\end{array} \right. ~~,\nn\\
\tilde{\alpha}_{33} &\simeq&\left\{
\begin{array}{l}  3.33 + 0.05 \tilde{\alpha}_{1\phi}
+
0.21 \tilde{\alpha}_{2\phi}-0.02 \tilde{\alpha}_{\Sigma}+ \cdots
\\3.40 + 0.05 \tilde{\alpha}_{1\phi}
-1.63 \tilde{\alpha}_{2\phi}- 0.001 \tilde{\alpha}_{G}+
\cdots \end{array} \right. ~~,\nn\\
\tilde{\alpha}_{3\phi} &\simeq&
\left\{
\begin{array}{l}  1.43 -0.58 \tilde{\alpha}_{1\phi}
-
1.43 \tilde{\alpha}_{2\phi}-0.03 \tilde{\alpha}_{\Sigma}+
\cdots\\
 0.88 -0.48 \tilde{\alpha}_{1\phi}
+8.83 \tilde{\alpha}_{2\phi}+ 0.01 \tilde{\alpha}_{G}+
\cdots\end{array} \right. ~~,\nn\\
\tilde{\alpha}_{H} &\simeq& \left\{
\begin{array}{l}
 1.08 -0.03 \tilde{\alpha}_{1\phi}
+0.10 \tilde{\alpha}_{2\phi}- 0.07 \tilde{\alpha}_{\Sigma}+
\cdots\\
2.51 -0.04 \tilde{\alpha}_{1\phi}
-1.68 \tilde{\alpha}_{2\phi}- 0.12 \tilde{\alpha}_{G}+
\cdots\end{array} \right. ~~,~~\\
\tilde{\alpha}_{\Sigma} &\simeq& \left\{
\begin{array}{l}
---\\
0.40 +0.01 \tilde{\alpha}_{1\phi}
-0.45 \tilde{\alpha}_{2\phi}-0.10 \tilde{\alpha}_{G}+
\cdots \end{array} \right. ~,\nn\\
\tilde{\alpha}_{\Sigma '} &\simeq& \left\{
\begin{array}{ll}
4.91 - 0.001 \tilde{\alpha}_{1\phi}
-0.03 \tilde{\alpha}_{2\phi}- 0.46 \tilde{\alpha}_{\Sigma}+
\cdots
\\8.30 + 0.01 \tilde{\alpha}_{1\phi}
+1.72 \tilde{\alpha}_{2\phi}- 0.36 \tilde{\alpha}_{G}+
\cdots \end{array} \right. ~~,  \nn\\
\tilde{\alpha}_{G} &\simeq& \left\{
\begin{array}{ll}
5.59 + 0.02 \tilde{\alpha}_{1\phi}
-0.04 \tilde{\alpha}_{2\phi}- 1.33 \tilde{\alpha}_{\Sigma}+
\cdots
\\--- \end{array} \right.   ~ ~.\nn
\ee
We have assumed that the Yukawa couplings $g_{IJ}$, except
$g_{33}$, vanish. They can be included into RGI relations
as small
perturbations,
but their numerical effects
will be rather small.

So far
we have assumed that
supersymmetry is unbroken. But we would like to recall
 that the RGI relations (13) and (14)
we have obtained above, remain unaffected by dimensional parameters
in mass-independent renormalization schemes.
Therefore, in the case of the soft breaking of supersymmetry,
these RGI relations are still valid.
We then have to translate the RGI
relations (13) and (14) into observable quantities.
To this end, we apply the
renormalization group technique and regard the RGI relations
 as
 the boundary conditions  holding
at the unification scale $M_{GUT}$ in addition to the
group theoretic one
$\alpha_{33}=\alpha_{t} ~=~\alpha_{b} ~=~\alpha_{\tau}$.
One of the large theoretical uncertainties in predicting
low energy parameters
is the arbitrariness of the superpartner
masses. To simplify our numerical analysis  we would like to assume a unique
threshold $M_{SUSY}$ for all the  superpartners.
Another arbitrariness is the number of the light Higgs particles
that are contained in $h_{i_R i_L}$ and also in
$G^{\mu}_{\nu~i_R i_L}$. The number $ N_{H}$ of the Higgses lighter
than $M_{SUSY}$ could vary from one to four while the number of
those to be taken into account above $M_{SUSY}$ is fixed at four.
In the following, we assume that $N_{H}=1$ and examine numerically the
evolution of the gauge and Yukawa couplings including the two-loop
 effects.

In table 1 we
present the low energy predictions of the present model for three
distinct  boundary conditions; $\tilde{\alpha}_{33}(M_{GUT})=4.0~,~3.2$
and
 $2.8$.
All the dimensionless parameters
(except $\tan \beta$) are defined in the $\overline{\rm MS}$
scheme, and all the masses (except for $M_{GUT}$ and $M_{SUSY}$)
are pole masses.

\vspace{0.2cm}
\noindent
\begin{tabular}{|c|c|c|c|c|c|c|c|}
\hline
$M_{ SUSY}$ [TeV]& $\tilde{\alpha}_{33}(M_{GUT})$   &$\alpha_{S}(M_{Z})$ &
$\alpha (M_{ GUT})$ &
 $\tan \beta$  &  $M_{GUT}$ [GeV]  & $m_{b}$ [GeV]& $m_{t} [GeV]$
\\ \hline
$1.6$ &4.0 &$0.119$ & $0.046$ & $63.0$ &
$0.9\times 10^{15}$ & $5.01$ & $197.8$
\\ \hline
$1.6$  &$3.2$  & $0.119$ & $0.046$ & $63.0$ & $ 0.9\times 10^{15}$
 & $4.97$
 & $196.1$  \\ \hline
$1.6$  &$2.8$  & $0.119$ & $0.046$ & $63.0$ & $ 0.9\times 10^{15}$
 & $4.95$
 & $195.1$  \\ \hline
\end{tabular}
\begin{center}
{\bf Table 1}. The predictions
for different boundary conditions, where we have used:\\
$m_{\tau}=1.78$ GeV,
$\alpha_{em}^{-1}(M_{Z})=127.9$ and $\sin\theta_{W}(M_{Z})
=0.2303$.\end{center}

\noindent
Note that the corrections
to $ \sin^{2} \theta_{W}(M_{Z})$ that come from a
large $m_{t}$, i.e., $ \sin^{2} \theta_{W}(M_{Z})=0.2324
-10^{-7}\, [138^2-(m_{t}/{\rm GeV})^2]$, are taken into account
 above. All the quantities in table 1, except for $M_{SUSY}$, are
predicted; the range of $\tilde{\alpha}_{33}$ is also given by the model
(see eq. (14)). We see from table that the low energy predictions
are insensitive against the value
of $\tilde{\alpha}_{33}$ and moreover that the predicted values
are consistent with the recent data of CDF and D$0$ \cite{cdf}.
 Another numerical analysis \cite{kubo3} shows that the present model
rather prefers large values of $M_{SUSY}$ ( $ > 400$ GeV).

\section{Conclusion}
The well-known unification attempts \cite{georgi1,fritzsch1} assume
that all the gauge interactions are unified at a
certain energy scale beyond which
they are described by a unified gauge theory based on a simple
gauge group.
The
measurements of the gauge couplings at LEP in fact suggest that the
minimal supersymmetric $SU(5)$ GUT \cite{dgs} is very
successful when comparing its theoretical values with
the experiments.
The GUTs  can also relate  Yukawa couplings among themselves, but
the GUT idea
alone cannot provide us with the possibility of relating the gauge and
Yukawa couplings.  In contrast to the GUT scheme, in the
alternative unification presented here the symmetry principles do
not play a mandatory r{\^ o}le.
The fundamental new concept here is to determine
renormalization
group invariant relations among couplings of a given GUT which
are valid before the breaking of the unifying gauge
symmetry. In the
particular application discussed here we have found successful GYU
yielding predictions which are consistent with the old and recent
experimental data.  Therefore, it is justified to hope that the
present unification scheme might be able to clarify further the
origin of the complex structure of the Yukawa sector of the standard
model.

\vspace{3cm}
\noindent
 {\em Acknowledgments}

\noindent
Part of the results presented here was obtained in the collaboration
with N.D. Tracas, and we thank him for the collaboration.
We would like to thank R. Oehme, K. Sibold and W. Zimmermann for
useful discussions and comments.
One of us (J.K) would like to thank the organizers of the workshop
for their warm hospitality.

\end{document}